\documentclass[10pt,twocolumn,a4paper]{ieeetran} 
\usepackage{times}
\usepackage{fullpage}
\usepackage{graphicx}

\newcommand{\figuramedia}[3]
{
\begin{figure*}
  \centering
 \includegraphics[width=13cm]{#1}
  \caption{#2}\label{#3}
\end{figure*}
}
\newcommand{\figurapiccola}[3]
{
\begin{figure}
  \centering
 \includegraphics[width=8cm]{#1} %
  \caption{#2}\label{#3}
\end{figure}
}
\newcommand{\figura}[3]
{
\begin{figure}
  \centering
 \includegraphics[width=7.5cm]{#1} %
  \caption{#2}\label{#3}
\end{figure}
}

\newcommand{\figuragrossa}[3]
{
\begin{figure}
  \centering
 \includegraphics[width=7.5cm]{#1}
  \caption{#2}\label{#3}
\end{figure}
}
\begin{document}
\title{On the Polyphase Decomposition for Design of Generalized Comb Decimation Filters
\thanks{This work was partially
supported by EuroConcepts., S.r.l. (www.euroconcepts.it), and by
Research Funds of MURST (Ministero dell'Universit\'{a} per la
Ricerca Scientifica Tecnologica).}}
\author{Massimiliano
Laddomada,~\IEEEmembership{Member,~IEEE}\\
\thanks{The author is with the Dipartimento di Elettronica, Politecnico di Torino, Corso Duca
degli Abruzzi 24, 10129 Torino, Italy.
E-mail:{\tt~laddomada@polito.it}}}
\maketitle
\begin{abstract}
Generalized comb filters (GCFs) are efficient anti-aliasing
decimation filters with improved selectivity and quantization
noise (QN) rejection performance around the so called folding
bands with respect to classical comb filters.

In this paper, we address the design of GCF filters by proposing
an efficient partial polyphase architecture with the aim to reduce
the data rate as much as possible after the $\Sigma \Delta $ A/D
conversion. We propose a mathematical framework in order to
completely characterize the dependence of the frequency response
of GCFs on the quantization of the multipliers embedded in the
proposed filter architecture. This analysis paves the way to the
design of multiplier-less decimation architectures.

We also derive the impulse response of a sample $3$rd order GCF
filter used as a reference scheme throughout the paper.
\end{abstract}
\section{Introduction and problem formulation}
The design of computationally efficient decimation filters for
$\Sigma \Delta$ A/D converters is a well-known research
topic~\cite{Temes,Hoge,candy_decim}. Given a base-band analog
input signal $x(t)$ with bandwidth $\left[-f_x,+f_x\right]$, a
$\Sigma \Delta$ A/D converter of order $B$ produces a digital
signal $x(nT_s)$ by sampling $x(t)$ at rate $f_s=2\rho f_x\gg
2f_x$, whereby $\rho$ is the so called oversampling ratio. The
normalized maximum frequency contained in the input signal is
defined as $f_c=\frac{f_x}{f_s}=\frac{1}{2\rho}$, and the digital
signal $x(nT_s)$ at the input of the first decimation filter has
frequency components belonging to the range $f_d\in [-f_c,f_c]$
(where $f_d$ denotes the digital frequency). This setup is
pictorially depicted in the reference architecture shown in
Fig.~\ref{arch}.
\figura{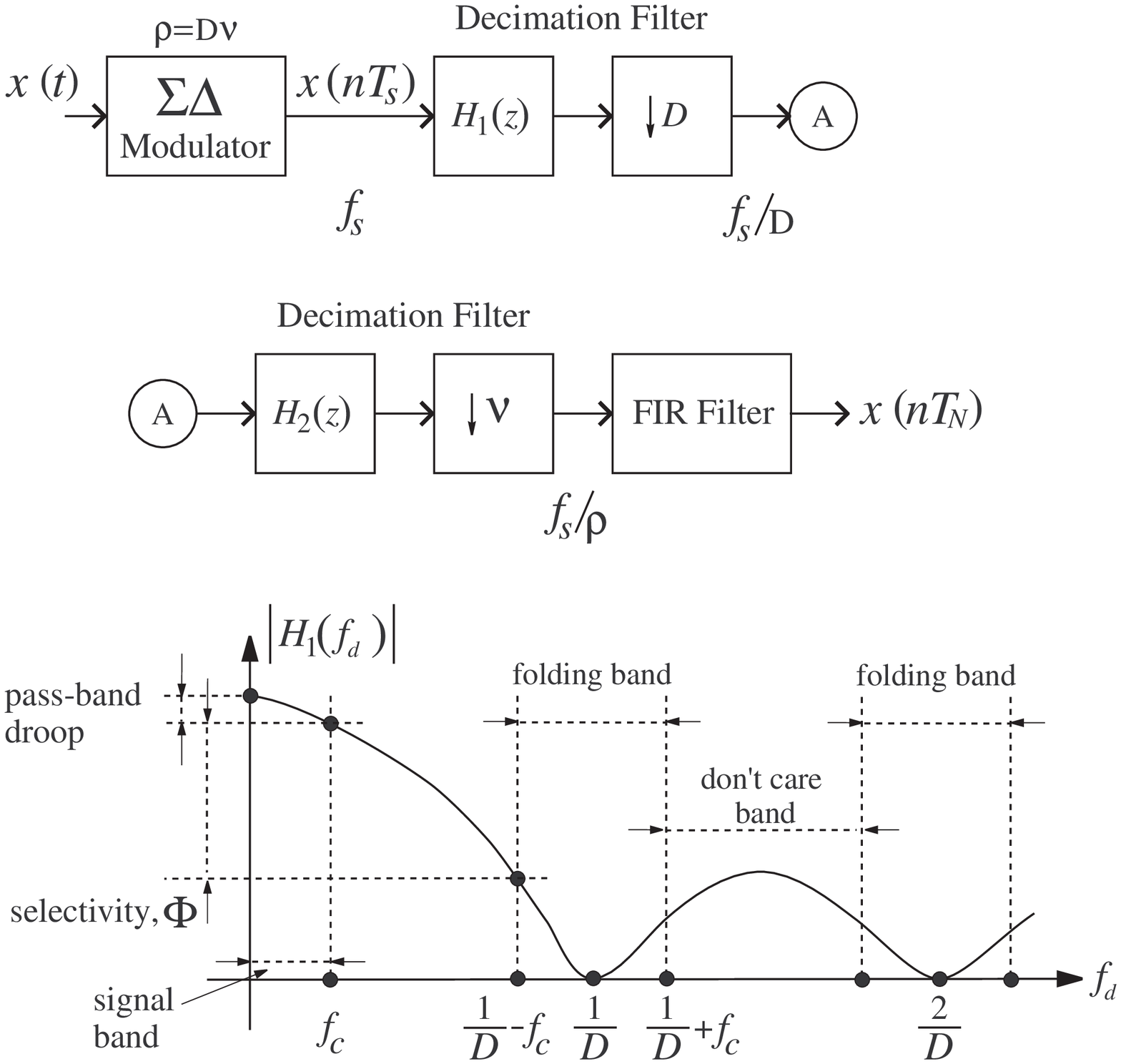}{General architecture of a
two-stage decimation chain for $\Sigma\Delta$ A/D converters,
along with a pictorial representation of the key frequency
intervals to be carefully considered for the design of the first
decimation stage.}{arch}

Owing to the condition $\rho\gg 1$, the decimation of an
oversampled signal $x(nT_s)$ is
efficiently~\cite{CrochiereRabiner} accomplished by cascading two
(or more) decimation stages, as highlighted in Fig.~\ref{arch},
followed by a FIR filter which provides the required selectivity
on the sampled signal $x(nT_N)$ at baseband. The first decimation
filter is usually an $N$-th order comb filter decimating by
$D$~\cite{Hoge,candy_decim,Chu}, whereby the order $N$ has to be
greater or equal to $B+1$~\cite{Temes,candy_decim}.

The design of a multistage decimation filter for $\Sigma\Delta$
converters poses stringent constraints on the shape of the
frequency response of the first decimation stage. Considering the
scheme in Fig.~\ref{arch}, the frequency response
$H_1(e^{j\omega})$ of the first decimation filter must attenuate
the QN falling inside the so called folding bands, i.e. the
frequency ranges defined as
 $\left[\frac{k}{D}-f_c;\frac{k}{D}+f_c\right]$ with $k=1,...,\lfloor
\frac{D}{2}\rfloor$ if $D$ is even, and $k=1,...,\lfloor
\frac{D-1}{2}\rfloor$ for $D$ odd (for conciseness, the set of
values assumed by $k$ will be denoted as $K_k$ throughout the
paper). The reason is that the $\Sigma\Delta$ QN falling inside
these frequency bands, will fold down to baseband because of the
sampling rate reduction by $D$ in the first decimation stage,
irremediably affecting the signal resolution after the multistage
decimation chain. This issue is especially important for the first
decimation stage, since the QN folding down to baseband has not
been previously attenuated. Fig.~\ref{arch} also shows the
frequency range $\left[0,f_c\right]$ where the useful signal
bandwidth falls, along with the so called \textit{don't care}
bands, i.e. the frequency ranges whose QN will be rejected by the
filters placed beyond the first decimation stage.

In connection with the first decimation stage in the multistage
architecture shown in Fig.~\ref{arch}, the required aliasing
protection around the folding bands is usually guaranteed by a
comb filter, which provides an inherent antialising function by
placing its zeros in the middle of each folding band. We recall
that the transfer function of a $N$-th-order comb filter is
defined as \cite{Hoge}:
\begin{equation}\small
H_{C_N}(z)=\left(\frac{1}{D}\frac{1-z^{-D}}{1-z^{-1}}\right)^N=
\frac{1}{D^N}\prod_{i=1}^{D-1}\left(1-z^{-1}e^{j\frac{2\pi}{D}i}\right)^N
\label{transf_funct_CIC_N}
\end{equation}
where $D$ is the desired decimation factor.

With this background, let us provide a quick survey of the recent
literature related to the problem addressed here. This survey is
by no means exhaustive and is meant to simply provide a sampling
of the literature in this fertile area.

A $3$rd order modified decimation sinc filter was proposed
in~\cite{Letizia}, and still further analyzed
in~\cite{max,Letizia_adnan}. The class of comb filters was then
generalized in~\cite{laddomada_gcf}, whereby the authors proposed
an optimization framework for deriving the optimal zero rotations
of GCFs for any filter order and decimation factor $D$.

Other works somewhat related to the topic addressed in this paper
are~\cite{gao}-\cite{laddomada_sharp}. In~\cite{gao}
and~\cite{aboushady} authors proposed computational efficient
decimation filter architectures for implementing classical comb
filters. In~\cite{kwentus} authors proposed the use of decimation
sharpened filters embedding comb filters, whereas
in~\cite{dolecek} authors addressed the design of a novel
two-stage sharpened comb decimator. In~\cite{laddomada}, authors
proposed novel decimation schemes for $\Sigma\Delta$ A/D
converters based on Kaiser and Hamming sharpened filters, then
generalized in \cite{laddomada_sharp} for higher order decimation
filters.

The main aim of this paper is to propose a flexible, yet
effective, partial polyphase architecture for implementing the GCF
filters proposed in the companion paper~\cite{laddomada_gcf}. To
this end, we first recall the $z$-transfer function of GCF filters
for completeness, and, then, provide a mathematical formulation
for deriving the impulse response of this class of decimation
filters. The latter is needed for deriving the polyphase
components of the proposed filters. In the second part of the
paper, the focus is on the sensitivity of the frequency response
of GCF filters due to the quantization of the multipliers embedded
in the proposed architecture. We also analyze zero displacements
in the $z$-transfer function of GCF filters for deducing useful
hints at the basis of any practical implementation of such
filters.

The rest of the paper is organized as follows. In Section~II, we
briefly recall the transfer functions of GCF filters and highlight
their main peculiarities with respect to classical comb filters.
Section~III presents an effective architecture, namely a partial
polyphase decomposition, for implementing GCF decimation filters;
the impulse response of a sample $3$rd order GCF filter is also
presented, and mathematically derived in the Appendix. In
Section~IV, we present a mathematical framework for evaluating the
sensitivity of the frequency response of GCF filters to the
approximations of the embedded multipliers. Zero displacements due
to multiplier approximation is discussed in Section~V, where we
also draw general guidelines for the design of such filters.
Finally, Section~VI draws the conclusions.
\section{The $z$-Transfer function of GCF filters}
\figurapiccola{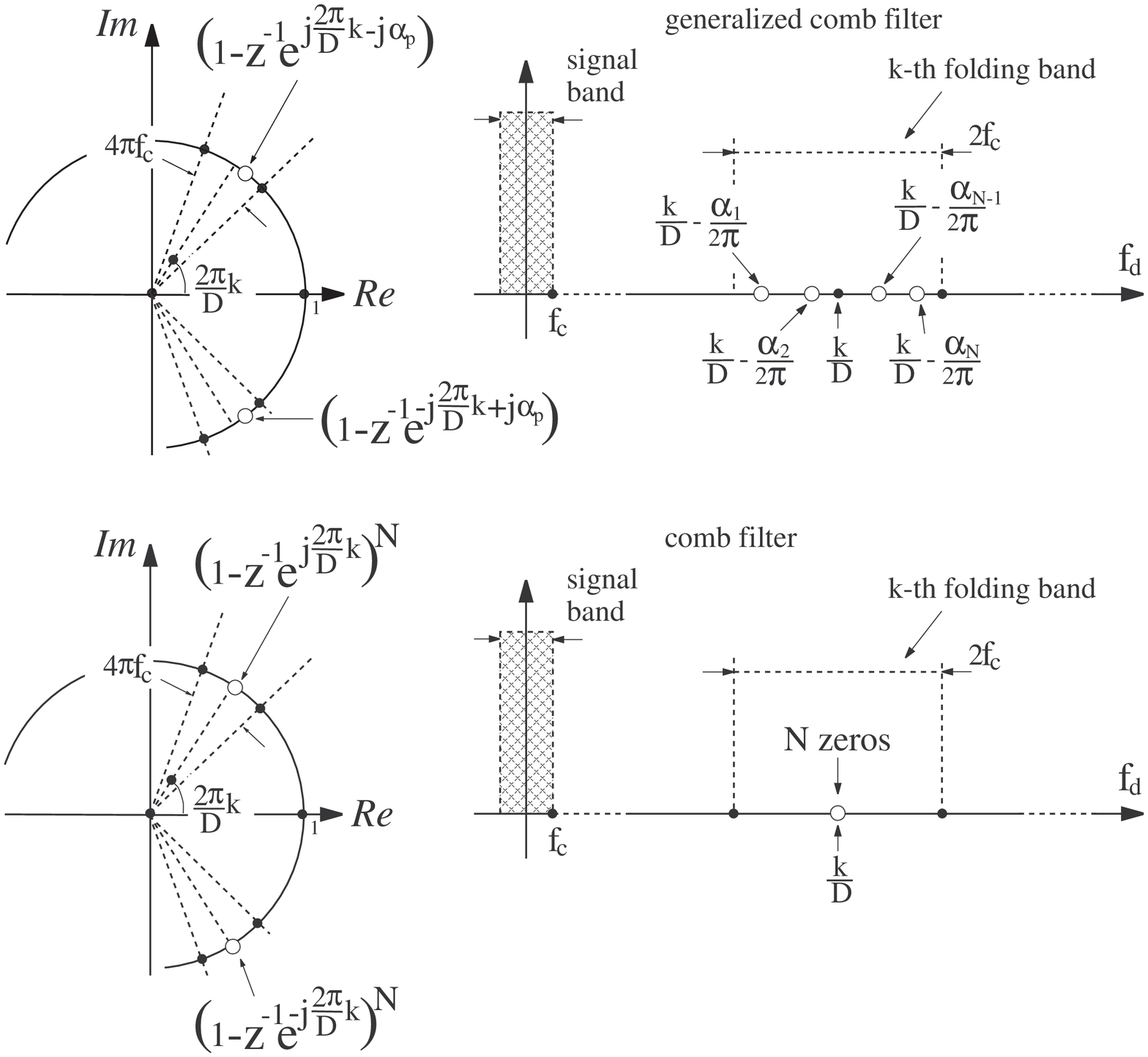}{Zero locations of the
considered decimation filters within the $k$-th folding band.
Zeros are displayed in the $z$-plane and in the frequency axis in
order to highlight their effect in both domains.}{zeros_posi_GCF}
In this section, we briefly recall the $z$-transfer function of
GCF filters proposed in~\cite{laddomada_gcf}. The $z$-transfer
function $H_{GCF_N}(z)$ of a $N$-th order GCF filter decimating by
$D$ is defined as:
\begin{eqnarray}\small\label{GCF}
\frac{1}{H_{o,ev,1}}H_1(z)&\cdot\prod_{n=1}^{\left\lfloor \frac{N}{2}\right\rfloor}H_2(z,\alpha_{n+N}),& D~even,~N~even\nonumber\\
\frac{\left(1+z^{-1}\right)}{H_{o,ev,2}}H_1(z)&\cdot\prod_{n=1}^{\left\lfloor \frac{N}{2}\right\rfloor}H_2(z,\alpha_{n+N}),& D~even,~N~odd\nonumber\\
\frac{1}{H_{o,od}}H_1(z),&& D~odd
\end{eqnarray}
whereby the involved basic functions are defined as follows:
\begin{equation}\small
\begin{array}{ll}
H_1(z)=\prod_{i=1}^{D_M}\prod_{n=1}^{N}\left(1-2\cos\left(\frac{2\pi}{D}i-\alpha_n\right)z^{-1}+z^{-2}\right),&\\
=\prod_{i=1}^{D_M}\prod_{n=1}^{N}\left(1-z^{-1}e^{+j\frac{2\pi}{D}i-j\alpha_n}\right)\left(1-z^{-1}e^{-j\frac{2\pi}{D}i+j\alpha_n}\right)&
\end{array}
\end{equation}
with $D_M= \frac{D}{2}-1$, for $D$ even, and $D_M= \frac{D-1}{2}$,
for $D$ odd, and
\begin{equation}
H_2(z,\alpha_{n+N})=1-2\cos\left(\pi-\alpha_{n+N}\right)z^{-1}+z^{-2}
\end{equation}
Terms $H_{o,ev,1}$, $H_{o,ev,2}$, and $H_{o,od}$ are appropriate
normalization constants chosen in such a way as to have
$\left.H_{GCF_N}(z)\right|_{z=1}=1$, and $\left\lfloor
\cdot\right\rfloor$ is the floor of the underlined number.

Let us summarize the main peculiarities of GCF filters by
comparing them to classical comb filters. GCFs are, as comb
filters, linear-phase filters since they are constituted by two
linear-phase basic filters, namely $H_1(z)$ and $H_2(z)$.

An $N$-th order comb filter~(\ref{transf_funct_CIC_N}) decimating
by $D$, places $N$-th order zeros in the complex locations
$z_i=e^{j\frac{2\pi}{D}i},~\forall i=1,\ldots,D-1$, or,
equivalently, in the digital frequencies
$f_{z_k}=\frac{k}{D},~k\in K_k$. On the other hand, an $N$-th
order GCF filter decimating by $D$ places $N$ pairs of conjugate
complex zeros in the $i$-th folding band\footnote{For conciseness,
we only deal with the positive semi-plane in the $z$-domain;
however, the zero placement is specular for what concerns the
zeros located in the lower semi-plane.}, with $i=1,\ldots,D_M$, as
exemplified by the function $H_1(z)$ in~(\ref{GCF}). A pictorial
representation of the zero locations of a GCF filter is given in
Fig.~\ref{zeros_posi_GCF}. The behaviour of the function $H_2(z)$
is analogous to that of $H_1(z)$ with the exception that its zeros
are placed around the location $z=-1$ in the $z$-complex plane,
and, it holds only for $D$ even. A convenient choice for
$\alpha_p$ is $ \alpha_p=q_p2\pi f_c,$~with $q_p\in
\left[-1,+1\right]$: this solution is such that each pair of
conjugate complex zeros falls inside the relative folding band
guaranteeing the required selectivity in these frequency bands. By
virtue of the zero distribution within the folding bands, GCF
filters provide improved $\Sigma\Delta$ QN rejection capabilities
with respect to classical comb filters of the same order. For
completeness, Table~\ref{optimal_zeros_positions_gcf} shows the
optimal zero rotations $q_p$s found in~\cite{laddomada_gcf} by
minimizing the $\Sigma\Delta$ QN around the folding bands. As an
example, a $3$rd order GCF filter provides a $\Sigma\Delta$ QN
rejection $8$dB higher than that guaranteed by a classical $3$rd
order comb filter. Throughout the paper we will use this optimal
choice of the zero rotations where no otherwise specified. We
invite the interested readers to refer to~\cite{laddomada_gcf} for
further details about the characteristics along with the
performance of GCF filters.
\begin{table}
\caption{Optimal parameters of GCF decimation filters.}
\begin{center}
\begin{tabular}{c|l|l|l|l} \hline

$N$ & $3$ & $4$ & $5$& $6$
\\\hline \hline

$q_1$& -0.79    & -0.35 & +0.55 & +0.95  \\\hline

$q_2$& 0.0      & +0.35 & +0.93 & +0.675 \\\hline

$q_3$& +0.79    &-0.88  & -0.55 & +0.25  \\\hline

$q_4$&+0.79     &+0.88  & -0.93 & -0.25 \\\hline

$q_5$&-         & +0.88 & 0.0 & -0.675\\\hline

$q_6$&-         &+0.35  &+0.55  & -0.95 \\\hline

$q_7$& -        & -     &+0.93  & +0.95 \\\hline

$q_8$&-         & -     & -     & +0.675\\\hline

$q_9$&-         & -     &-      & +0.25 \\\hline

$G-[dB]$& $\sim$8 &$\sim$13 & $\sim$18&$\sim$ 23\\\hline\hline

\end{tabular}
\label{optimal_zeros_positions_gcf}
\end{center}
\end{table}
\section{Partial Polyphase Decomposition of GCF Filters}
\label{non_recursive_impl}
\figuramedia{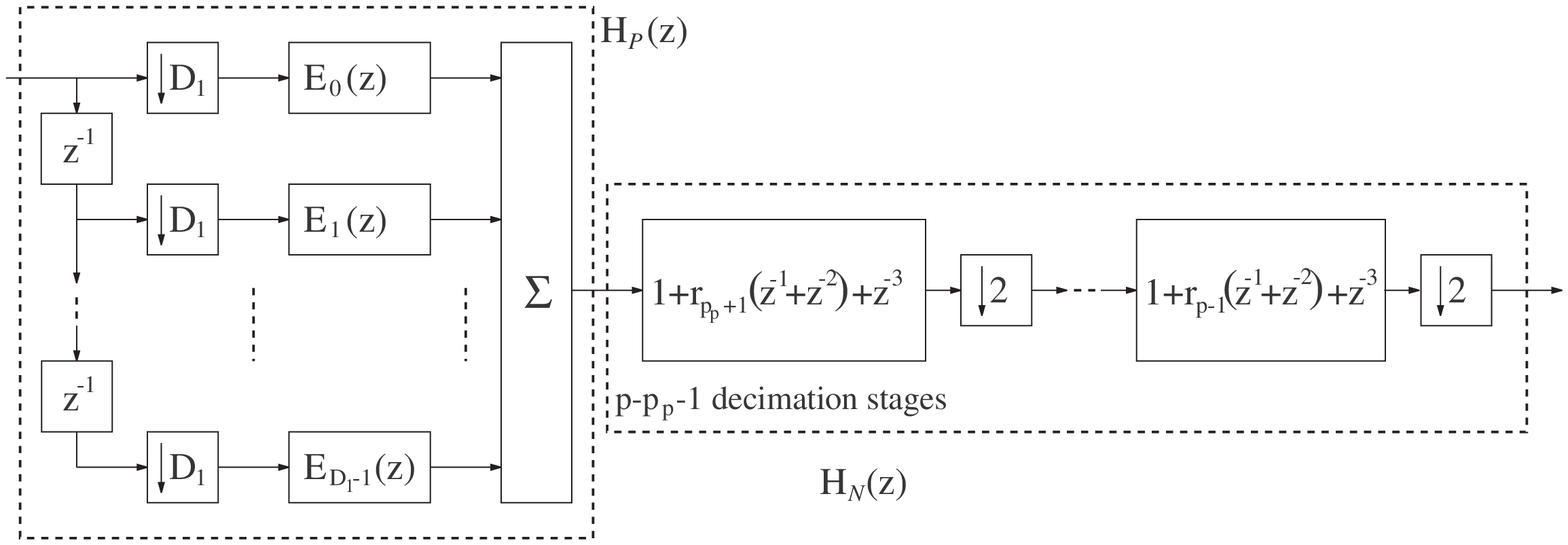}{Architecture of the
partial polyphase implementation of the decimation filter
$H_{GCF_3}(z)$. The overall decimation factor $D=D_1\cdot D_2$ is
split between the polyphase section decimating by $D_1=2^{p_p+1}$,
and the non recursive section, decimating by $D_2=2^{p-p_p-1}$,
composed of $p-p_p-1$ decimation stages each one decimating by
$2$. Integer $p_p$ can take on any value in the set
$\left[-1,p-1\right]$.}{n_rec_impl_gcf3}
This section presents a non-recursive, partial polyphase
implementation of GCF filters, suitable for decimation factors $D$
that can be expressed as the $p$-th power-of-two, i.e. $D=2^p$
with $p$ a suitable integer greater than zero. For conciseness, we
only address the implementation of a $3$rd-order GCF filter, but
the considerations that follow, can be easily extended to higher
order GCF filters with $D$ as specified above.

Let us focus on the optimal zero rotations shown in
Table~\ref{optimal_zeros_positions_gcf}, and consider the
$z$-transfer function $H_{GCF_3}(z)$ derived in~(\ref{GCF}). Due
to the symmetry of the $q_p$s in
Table~\ref{optimal_zeros_positions_gcf}, the zeros belonging to
$H_{GCF_3}(z)$ can be collected in the following three
$z$-transfer functions:
\begin{eqnarray}
\prod_{i=1}^{D-1}\left(1-z^{-1}e^{j\frac{2\pi}{D}i}\right)&=&\frac{1-z^{-D}}{1-z^{-1}}\nonumber\\
\prod_{i=1}^{D-1}\left(1-z^{-1}e^{j\frac{2\pi}{D}i}e^{-j\alpha}\right)&=&\prod_{i=1}^{D-1}\left(1-\beta_1^{-1}e^{j\frac{2\pi}{D}i}\right)\nonumber\\
\prod_{i=1}^{D-1}\left(1-z^{-1}e^{j\frac{2\pi}{D}i}e^{+j\alpha}\right)&=&\prod_{i=1}^{D-1}\left(1-\beta_2^{-1}e^{j\frac{2\pi}{D}i}\right)\nonumber
\end{eqnarray}
whereby $\beta_1=z\cdot e^{j\alpha},~\beta_2=z\cdot e^{-j\alpha}$,
and $\alpha=|q_1|2\pi f_c$. Notice that the first $z$-transfer
function accounts for the zeros falling in the digital frequencies
$\frac{k}{D},~k\in K_k$, which corresponds to the zeros of a
classical $1$st order comb filter. The other two $z$-transfer
functions consider the rotated zeros.

The $z$-transfer function $H_{GCF_3}(z)$ can be easily obtained by
multiplying the previous three $z$-transfer
functions\footnote{Notice that, for conciseness, in the
mathematical formulation that follows, we omit the constant term
assuring unity gain at base-band.}:
\begin{equation}\label{third_order_gcf3}\small
H_{GCF_3}(z)=\frac{1-z^{-D}}{1-z^{-1}}\frac{1-z^{-D}e^{j\alpha
D}}{1-z^{-1}e^{j\alpha}}\frac{1-z^{-D}e^{-j\alpha
D}}{1-z^{-1}e^{-j\alpha}}
\end{equation}
The impulse response of the $3$rd order GCF filter
in~(\ref{third_order_gcf3}) has been derived in the Appendix.

A non-recursive implementation of filter $H_{GCF_3}(z)$ can be
obtained by expressing each rational function
in~(\ref{third_order_gcf3}) in a non recursive form. By doing so,
the first polynomial ratio can be rewritten as follows:
\begin{equation}\label{formula_comb_1_ordine}
\frac{1-z^{-D}}{1-z^{-1}}=\sum_{i=0}^{D-1}z^{-i}=\prod_{i=0}^{\log_2(D)-1}\left(1+z^{-2^i}\right)
\end{equation}
whereby last equality holds for $D=2^p$. Upon using a similar
reasoning, it is straightforward to observe that the following
equality chain, which derives from~(\ref{formula_comb_1_ordine})
by imposing $\beta=z\cdot e^{\mp j\alpha}$, holds as well:
\begin{equation}\label{formula_comb_1_ordine_zr}
\begin{array}{ll}
\frac{1-z^{-D}e^{\pm j\alpha D}}{1-z^{-1}e^{\pm
j\alpha}}=\sum_{i=0}^{D-1}z^{-i}e^{\pm j i\alpha
}=&\\
=\prod_{i=0}^{\log_2(D)-1}\left(1+z^{-2^i}e^{\pm j 2^i\alpha
}\right)&
\end{array}
\end{equation}

\noindent By noting that
\[
\begin{array}{ll}
\left(1+z^{-2^i}e^{+j2^i\alpha}\right)\left(1+z^{-2^i}e^{-j2^i\alpha}\right)=&\\
=1+2\cos(2^i\alpha)z^{-2^i}+z^{-2^{i+1}}
\end{array}
\]
after some algebra,~(\ref{third_order_gcf3}) can be rewritten as
follows:
\begin{equation}\label{third_order_gcf3_2}
\begin{array}{ll}
H_{GCF_3}(z) =
\prod_{i=0}^{\log_2(D)-1}\left[\left(1+z^{-2^i}\right)\cdot \right.&\\
\left.\cdot\left(1+2\cos(2^i\alpha)z^{-2^i}+z^{-2^{i+1}}\right)\right]=&\\
= \prod_{i=0}^{\log_2(D)-1}\left[ 1+r_i\cdot\left( z^{-2^i}+
z^{-2\cdot 2^i}\right)+z^{-3\cdot 2^{i}} \right]&
\end{array}
\end{equation}
whereby
\[
\begin{array}{ll}
r_i=1+2\cos\left(2^i\alpha\right)=1+2\cos\left(
q\frac{2^i\pi}{\rho}\right)=&\\
=1+2\cos\left( q2^{i+1}\pi f_c\right),~\forall
i=0,\ldots,\log_2(D)-1 &
\end{array}
\]
Assume that the decimation factor $D$ can be decomposed as follows
$D=D_1\cdot D_2$, whereby $D_1=2^{p_p+1}$ and $D_2=2^{p-p_p-1}$.
By doing so,~(\ref{third_order_gcf3_2}) can be rewritten as
follows:
\begin{eqnarray}\label{third_order_gcf3_2_2}
H_{GCF_3}(z) &=& H_{P}(z) \cdot H_{N}(z)
\end{eqnarray}
whereby
\begin{equation}\label{third_order_gcf3_2_3}
\begin{array}{ll}
H_{P}(z) =\prod_{i=0}^{p_p}\left[ 1+r_i\cdot\left( z^{-2^i}+
z^{-2\cdot 2^i}\right)+z^{-3\cdot 2^{i}} \right]&\\
H_{N}(z) =\prod_{i=p_p+1}^{p-1}\left[ 1+r_i\cdot\left( z^{-2^i}+
z^{-2\cdot 2^i}\right)+z^{-3\cdot 2^{i}} \right]&\\
\end{array}
\end{equation}
Remembering that $p_p=\log_2 (D_1)-1$, it is straightforward to
observe that $H_{P}(z)$ is the $z$-transfer function of a $3$rd
order GCF filter decimating by $D_1$. The impulse response
$h_{P}(n),\forall n\in [0,3D_1-3],$ of filter $H_{P}(z)$ has been
derived in Appendix:
\begin{eqnarray}\label{inv_z_trans_6_copia}
h_{P}(n)=e^{+j\alpha n}\sum_{k_3=0}^{n}e^{-2j\alpha
k_3}\sum_{k_2=0}^{k_3}e^{j\alpha k_2} \sum_{k_1=0}^{k_2}x_t(k_1)&
\end{eqnarray}
The sequence $x_t(n)$ is defined as follows:
\begin{equation}\label{inv_z_trans_X_tz_copia}
x_t(n)= \delta(n)-r\delta(n-D_1)+r\delta(n-2D_1)-\delta(n-3D_1)
\end{equation}
whereby $r=1+\cos (\alpha D_1)$, and $\alpha=|q_1|\pi
\frac{1}{\rho}$.

The polyphase decomposition of the $z$-transfer function
$H_{P}(z)$ is defined as follows:
\begin{equation}\label{third_order_gcf3_2_poly_dec}
\begin{array}{ll}
H_{P}(z) =\sum_{k=0}^{D_1-1}z^{-k}E_k\left(z^{D_1}\right)
\end{array}
\end{equation}
whereby the functions $E_k(z)$ are the polyphase components.
Time-domain coefficients related to $E_k(z)$ are defined as
follows:
\begin{equation}\label{polifase_generale_2}
e_k(n)=h_{P}(D_1 n+k),~\forall k\in [0,D_1-1]
\end{equation}
and can be easily obtained by employing the recursive equation
in~(\ref{inv_z_trans_6_copia}).

Some observations are in order. By choosing $p_p=p-1$, GCF filter
$H_{GCF_3}(z)$ is fully realized in polyphase form, whereas for
$p_p=-1$ the overall decimator is realized as the cascade of $p$
non recursive decimation stages each one decimating by $2$. Any
other value of $p_p\in [0,p-2]$ yields a partial polyphase
decomposition.

The natural question that arises at this point concerns the
practical implementation of the GCF filter $H_{GCF_3}(z)$. In the
following we derive an architecture for implementing GCF filters,
while in the next section we present a mathematical framework for
highlighting the sensitivity of the proposed architecture to the
approximation of its multipliers. The latter is needed for
deducing useful hints at the basis of multiplier-less
implementations of the proposed filters.

By applying the commutative property employed in \cite{Chu}, it is
possible to obtain the cascaded architecture shown in
Fig.~\ref{n_rec_impl_gcf3}. The first polyphase decimation stage
allows the reduction of the sampling rate by $D_1$, thus reducing
the operating rate of the subsequent decimation stages belonging
to $H_N(z)$. Any stage of $H_N(z)$ in Fig.~\ref{n_rec_impl_gcf3}
is constituted by a simple FIR filter operating at a different
rate. Such an example, the $i$-th stage, with $i\in [0,p-p_p-2]$,
is characterized by the transfer function
$\left[1+r_i\left(z^{-1}+z^{-2}\right)+z^{-3}\right]$ operating at
rate $f_s/\left(D_1\cdot 2^{i}\right)$ ($f_s$ is the
$\Sigma\Delta$ sampling frequency).

The frequency response related to $H_N(z)$ can be evaluated by
substituting $z=e^{j\omega}$ in~(\ref{third_order_gcf3_2_3}):
\begin{equation}\label{third_order_gcf_fr}
\begin{array}{ll}
H_{N}(e^{j\omega}) = 2 \prod_{u=p_p+1}^{p-1}e^{-j3\cdot
2^{u-1}\omega}\cdot&\\
\cdot\left[ \cos\left(3\cdot 2^{u-1}\omega\right) +r_u \cos\left(
2^{u-1}\omega\right) \right]
\end{array}
\end{equation}
whereby $\omega=2\pi f_d$, and $r_u$ is defined as:
\begin{equation}\label{moltiplicatori_ru}
r_u=1+2\cos\left(2^u\alpha\right), ~\forall u=p_p+1,\ldots,p-1
\end{equation}
\section{Sensitivity Analysis}
This section deals with the analysis of the sensitivity of filter
$H_{GCF_3}(z)$ to the approximation of its multipliers. In brief,
the goal is to design the non-recursive architecture in
Fig.~\ref{n_rec_impl_gcf3} without multipliers, while guaranteeing
the gain (8~dB based on the results shown in
Table~\ref{optimal_zeros_positions_gcf}) in terms of
$\Sigma\Delta$ QN rejection with respect to a classical comb
filter.

Let us evaluate the sensitivity of the frequency response
in~(\ref{third_order_gcf3_2_2}) with respect to its coefficients.
Notice that there are two sets of coefficients: $L=3D_1-2$
multipliers (i.e. $c_{n,k}=h_P(D_1 n+k)$) belong to the polyphase
section $H_P(z)$, and $p-p_p-1$ multipliers $r_u$ (shown
in~(\ref{moltiplicatori_ru})) belong to the decimation filter
$H_N(z)$.

First of all, notice that when the generic coefficient $c_{n,k}$
($r_u$) is approximated, its actual value can be expressed as
$\widetilde{c}_{n,k}=c_{n,k}+\Delta
c_{n,k}~(\widetilde{r}_{u}=r_u+\Delta r_u)$, whereby $\Delta
c_{n,k}$ ($\Delta r_u$) is the approximation error. On the other
hand, the approximations of coefficients $c_{n,k}$ and $r_u$ imply
that $H_{GCF_3}(e^{j\omega})$ be written as
\begin{equation}\label{hgcf3_diff}
\widetilde{H}_{GCF_3}(e^{j\omega})=H_{GCF_3}(e^{j\omega})+\Delta
H_{GCF_3}(e^{j\omega})
\end{equation}
The dependence of the frequency response $H_{GCF_3}(e^{j\omega})$
on the approximation of its multipliers can be evaluated by
differentiating~(\ref{third_order_gcf3_2_2}):
\begin{equation}\small\label{sensitivity_hgcf_3_1}
\Delta H_{GCF_3}(e^{j\omega})=H_{N}(e^{j\omega})\Delta
H_{P}(e^{j\omega})+H_{P}(e^{j\omega})\Delta H_{N}(e^{j\omega})
\end{equation}
whereby
\begin{equation}\small\label{sensitivity_delta_HN}
H_{P}(e^{j\omega})\Delta
H_{N}(e^{j\omega})=H_{P}(e^{j\omega})\sum_{u=p_p+1}^{p-1}
\frac{\partial H_{N}(e^{j\omega}) }{\partial r_u} \Delta r_u
\end{equation}
and
\begin{equation}\small\label{sensitivity_delta_HP}
H_{N}(e^{j\omega})\Delta
H_{P}(e^{j\omega})=H_{N}(e^{j\omega})\left[\sum_{k,n}\frac{\partial
H_{P}(e^{j\omega}) }{\partial c_{n,k}}\Delta c_{n,k}\right]
\end{equation}
with
\begin{equation}\label{F_k_n}
H_{P}(e^{j\omega})=\sum_{k=0}^{D_1-1}\sum_{n=0}^{\left\lfloor
L/D_1\right\rfloor} c_{n,k} e^{-j\omega (D_1 n+ k)}
\end{equation}
\noindent Let us evaluate the derivative of $H_{N}(e^{j\omega})$
with respect to $r_u,~\forall u=p_p+1,\ldots,p-1$:
\begin{equation}\label{sensitivity_hgcf_3_2}
\begin{array}{ll}
\frac{\partial H_{N}(e^{j\omega}) }{\partial r_u}=2e^{-j3\cdot
2^{u-1}\omega}\cos\left(
2^{u-1}\omega\right)\prod_{m=p_p+1,~m\ne u}^{p-1}&\\
e^{-j3\cdot 2^{m-1}\omega} \left[ \cos\left(3\cdot
2^{m-1}\omega\right) +r_m\cdot \cos\left( 2^{m-1}\omega\right)
\right]&
\end{array}
\end{equation}
Equation~(\ref{sensitivity_hgcf_3_2}) can be rewritten as follows:
\begin{equation}\label{sensitivity_hgcf_3_2_2}
\begin{array}{ll}
\frac{\partial H_{N}(e^{j\omega}) }{\partial
r_u}=H_{N}(e^{j\omega})\cdot\frac{\cos\left(2^{u-1}\omega\right)
}{\cos\left(3\cdot 2^{u-1}\omega\right)+r_u\cdot \cos\left(
2^{u-1}\omega\right)}&
\end{array}
\end{equation}

Upon substituting~(\ref{sensitivity_hgcf_3_2_2})
in~(\ref{sensitivity_delta_HN}), it is possible to obtain:
\begin{equation}\label{sensitivity_hgcf_3_3}
\begin{array}{ll}
H_{P}(e^{j\omega})\Delta H_{N}(e^{j\omega})= H_{P}(e^{j\omega})H_{N}(e^{j\omega})\cdot &\\
\cdot
\sum_{u=p_p+1}^{p-1}\frac{\cos\left(2^{u-1}\omega\right)\Delta r_u
}{\cos\left(3\cdot 2^{u-1}\omega\right)+r_u\cdot \cos\left(
2^{u-1}\omega\right)}=&\\
=H_{GCF_3}(e^{j\omega})\cdot\sum_{u=p_p+1}^{p-1}\frac{\cos\left(2^{u-1}\omega\right)\Delta
r_u }{\cos\left(3\cdot 2^{u-1}\omega\right)+r_u\cdot \cos\left(
2^{u-1}\omega\right)}&
\end{array}
\end{equation}

Let us consider $\Delta H_{P}(e^{j\omega})$. Given
$H_{P}(e^{j\omega})$ in~(\ref{F_k_n}), it is straightforward to
obtain the following relation:
\[
\frac{\partial H_{P}(e^{j\omega}) }{\partial c_{n,k}}=e^{-j\omega
(D_1 n+ k)}
\]
By substituting the previous equation in
(\ref{sensitivity_delta_HP}), it is possible to obtain:
\[
\Delta
H_{P}(e^{j\omega})=\sum_{k=0}^{D_1-1}\sum_{n=0}^{\left\lfloor
L/D_1\right\rfloor} \Delta c_{n,k} e^{-j\omega (D_1 n+ k)}
\]
Upon multiplying and dividing by $H_P(e^{j\omega})$, the function
$H_N(e^{j\omega})\Delta H_{P}(e^{j\omega})$ can be rewritten as:
%
%
\begin{equation}
\frac{H_{GCF_3}(e^{j\omega})}{H_P(e^{j\omega})}\sum_{k=0}^{D_1-1}\sum_{n=0}^{\left\lfloor
L/D_1\right\rfloor} \Delta c_{n,k} e^{-j\omega (D_1 n+ k)}
\end{equation}

The actual frequency response $\widetilde{H}_{GCF_3}(e^{j\omega})$
in~(\ref{hgcf3_diff}) can be expressed as follows:
\begin{equation}\small\label{sensitivity_hgcf_3_4}
\begin{array}{ll}
\widetilde{H}_{GCF_3}(e^{j\omega})=H_{GCF_3}(e^{j\omega})+\Delta
H_{GCF_3}(e^{j\omega})=&\\
=H_{GCF_3}(e^{j\omega})
\cdot\left[1+\frac{1}{H_P(e^{j\omega})}\sum_{k=0}^{D_1-1}\sum_{n=0}^{\left\lfloor
L/D_1\right\rfloor} \left[\Delta c_{n,k}\cdot \right.\right.&\\
\left.\left.\cdot e^{-j\omega (D_1 n+ k)}\right]+
\sum_{i=p_p+1}^{p-1}\frac{\cos\left(2^{i-1}\omega\right) \Delta
r_i}{\cos\left(3\cdot 2^{i-1}\omega\right)+r_i\cdot \cos\left(
2^{i-1}\omega\right)} \right]&
\end{array}
\end{equation}

The effects of the approximation of the multipliers $c_{n,k}$ and
$r_u$ on the actual frequency response
$\widetilde{H}_{GCF_3}(e^{j\omega})$ can be understood by
analyzing the frequency behavior of the following error function:
\begin{equation}\label{sensitivity_hgcf_3_5}
\begin{array}{ll}
\Delta
H\left(e^{j\omega}\right)=\frac{1}{H_P(e^{j\omega})}\sum_{k=0}^{D_1-1}\sum_{n=0}^{\left\lfloor
L/D_1\right\rfloor} \left[\Delta c_{n,k}\cdot \right.&\\
\left.\cdot e^{-j\omega (D_1 n+ k)}\right]+
\sum_{i=p_p+1}^{p-1}\frac{\cos\left(2^{i-1}\omega\right) \Delta
r_i}{\cos\left(3\cdot 2^{i-1}\omega\right)+r_i\cdot \cos\left(
2^{i-1}\omega\right)} &\\
=\Delta H_1(e^{j\omega})+\sum_{i=p_p+1}^{p-1}\Delta
H_{2,i}(e^{j\omega})&
\end{array}
\end{equation}

\noindent which, to some extent, quantifies the distortion between
the desired $H_{GCF_3}(e^{j\omega})$ and the actual frequency
response $\widetilde{H}_{GCF_3}(e^{j\omega})$.

Fig.~\ref{sensitivity_D16_K4} depicts the frequency behaviours of
the error functions $\Delta H_1\left(e^{j\omega}\right)$ and
$\Delta H_{2,i}\left(e^{j\omega}\right)$ noted in the last row
of~(\ref{sensitivity_hgcf_3_5}), for the following sample set of
parameters: $D=32,~D_1=8,~D_2=4$, $p_p=2$, $p=5$, $\nu=4$,
$q=|q_1|=0.79$, and $\Delta h_P(n)=\Delta r_i= 10^{-4},~\forall
n,i$.

Fig.~\ref{sensitivity_D16_K4_single} shows the behaviours of the
frequency responses $H_P(e^{j\omega})$ and $H_N(e^{j\omega})$ for
the sample set of parameters noted in the respective label.
\figuragrossa{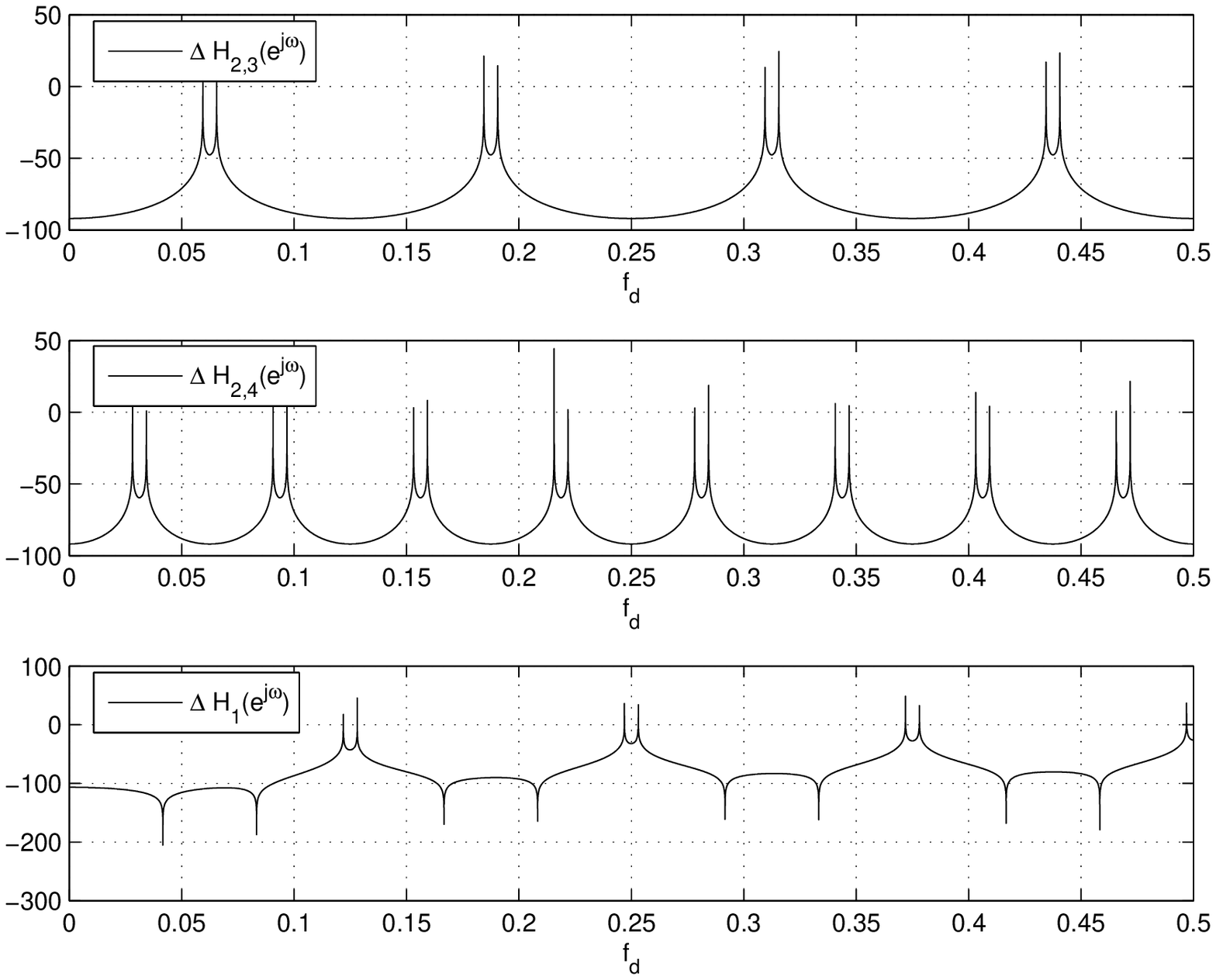}{Frequency behaviours
in dB of the functions $\left|\Delta
H_1\left(e^{j\omega}\right)\right|$, $\left|\Delta
H_{2,3}\left(e^{j\omega}\right)\right|$, and $\left|\Delta
H_{2,4}\left(e^{j\omega}\right)\right|$ in
(\ref{sensitivity_hgcf_3_5}) for $D=32,~D_1=8,~D_2=4$, $p_p=2$,
$p=5$, $\nu=4$, $q=0.79$, and $\Delta h_P(n)=\Delta r_i=
10^{-4},~\forall n,i$.}{sensitivity_D16_K4}

Some key observations are in order. Fig.~\ref{sensitivity_D16_K4}
shows that the pass-band behaviour of the filter
    $H_{GCF_3}(e^{j\omega})$ is not affected
    by the approximation of its coefficients. Sensitivity of
    $H_{GCF_3}(e^{j\omega})$ is very low
    for $f_d\in \left[0,f_c\right]$, whereby
    $f_c=\frac{1}{2\rho}$. This in turn suggests that the filter pass-band
    droop does not degrade by virtue of multipliers' approximations,
    and it is as low as the one guaranteed by filter
    $H_{GCF_3}(e^{j\omega})$. Notice also that the sensitivity is very
    low around the digital frequency $\frac{1}{D}-f_c$, which
    defines the selectivity of the decimation filter~\cite{Temes}.

Fig.s~\ref{sensitivity_D16_K4} and~\ref{sensitivity_D16_K4_single}
show that the approximation of multipliers $h_P(n)$ and $r_u$
affects the sensitivity of the frequency response
$H_{GCF_3}(e^{j\omega})$ in disjoint folding bands. Indeed,
filters $H_{P}(e^{j\omega})$ and $H_{N}(e^{j\omega})$ place the
respective zeros in different digital frequencies, as clearly
highlighted in Fig.~\ref{sensitivity_D16_K4_single}.

Fig.~\ref{sensitivity_D16_K4} also shows that the frequency
response $H_{GCF_3}(e^{j\omega})$
    is very
    sensitive to coefficients' approximations specially in the folding
    bands $\left[\frac{k}{D}-f_c,\frac{k}{D}+f_c\right],~\forall k\in
    K_k$. This in turn suggests that particular care must be devoted to
the approximation of multipliers embedded in both $H_P(z)$ and
$H_N(z)$ in order to preserve the QN rejection performance around
the folding bands. However, the same figure suggests that the
approximation of the multipliers $r_u$s belonging to $H_N(z)$, can
be done independently from the approximations of coefficients
$h_P(n)$ belonging to $H_P(z)$.

Sensitivity analysis derived above, allows to draw a general
picture of the effects of the approximations of the coefficients
on the actual frequency response $\widetilde{H}_{GCF_3}(z)$.
Nevertheless, we deduced that the sensitivity is very low in the
pass-band $[0,f_c]$ and around the frequency $\frac{1}{D}-f_c$.
This in turn suggests that both pass-band droop and selectivity of
GCF filters are preserved by the approximations of the
multipliers.

An important question is still open. We still need to quantify the
extent of the effects of the coefficient approximations on the
frequency response $\widetilde{H}_{GCF_3}(z)$. In brief, the basic
question we want to answer is as follows. What are the
approximation errors $\Delta r_u,~\forall u=p_p+1,\ldots,p-1,$ and
$\Delta h_P(n),~\forall n\in [0,3D_1-3],$ that we can tolerate on
$\widetilde{H}_{GCF_3}(z)$? The answer to this question is the
focus of the next section. It is anticipated that whatever the
condition on the maximum approximation error tolerated on the
frequency response $\widetilde{H}_{GCF_3}(e^{j\omega})$, it should
be related to the behaviour of such a function around the folding
bands, since both pass-band droop and selectivity of these filters
are mainly unaffected by the approximation of the multipliers.
\section{Estimation of zero displacements due to coefficient approximations}
\figuragrossa{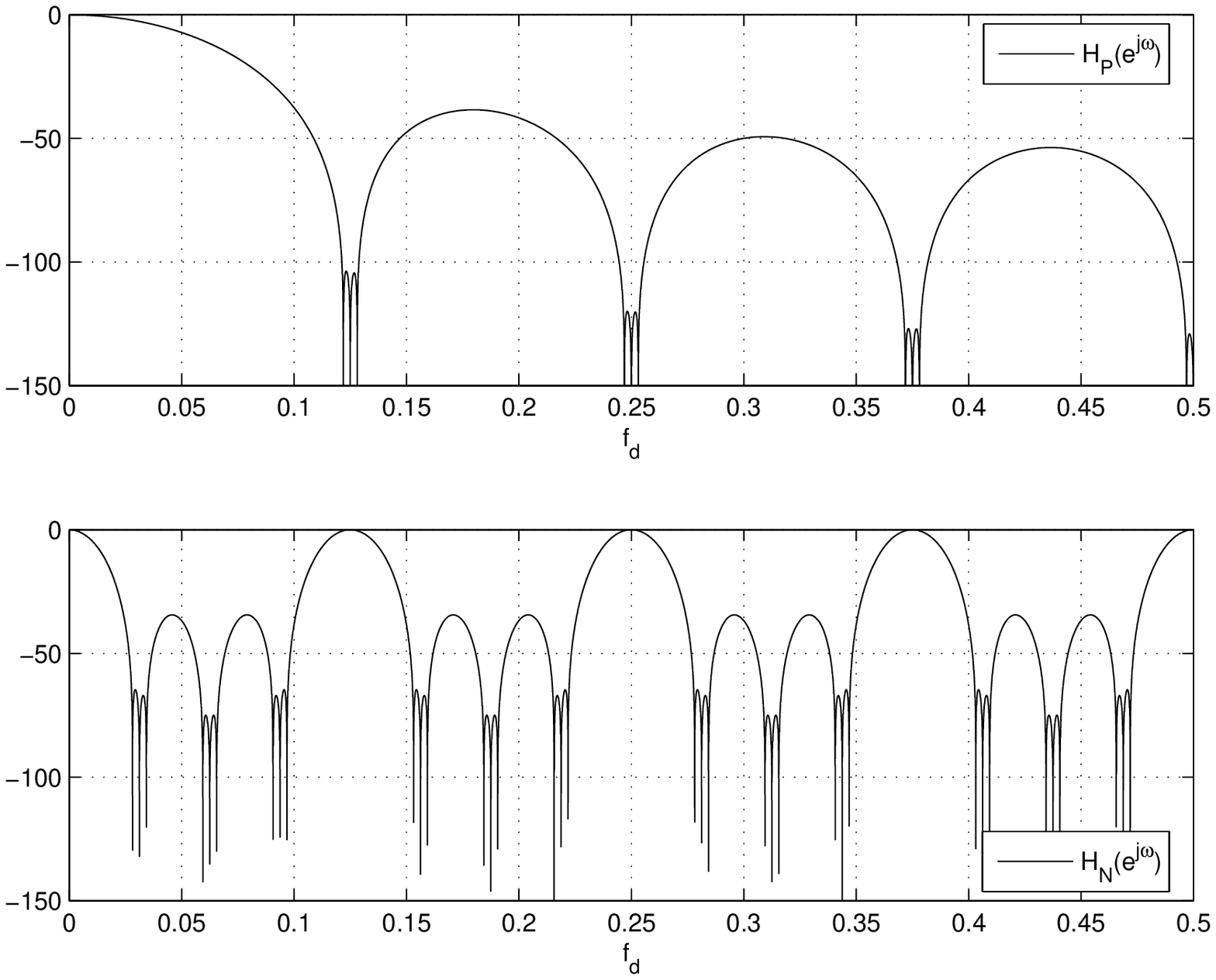}{Frequency
behaviours in dB of the functions
$\left|H_P\left(e^{j\omega}\right)\right|$ and
$\left|H_N\left(e^{j\omega}\right)\right|$, respectively,
in~(\ref{third_order_gcf3_2_poly_dec})
and~(\ref{third_order_gcf_fr}) for $D=32,~D_1=8,~D_2=4$, $p_p=2$,
$p=5$, $\nu=4$, $q=0.79$.}{sensitivity_D16_K4_single}
Besides improving the selectivity on the frequency
$\frac{1}{D}-f_c$, GCF filters provide improved $\Sigma\Delta$ QN
rejection around the folding bands with respect to classical,
equal-order comb filters~\cite{laddomada_gcf}. However,
coefficient approximations can have detrimental effects on the
zero locations in the $z$-plane, and, accordingly, can worsen
$\Sigma\Delta$ QN rejection performance around the folding bands.
As a consequence, it is useful to estimate the effects of
coefficient approximations on the actual $\Sigma\Delta$ QN
rejection performance guaranteed by filter
$\widetilde{H}_{GCF_3}(e^{j\omega})$ by estimating the induced
zero displacements. Next three sections address this topic by
first examining filters $H_P(z)$ and $H_N(z)$
separately\footnote{This is possible by virtue of the sensitivity
analysis derived above: zeros of both $H_P(z)$ and $H_N(z)$
affects different folding bands. In other words, each pair of
conjugate complex zeros affects the behaviour of the frequency
response in only one folding band.}, and then deducing some hints
from the derived theoretical analysis.
\subsection{Displacements of zeros belonging to $H_P(z)$}
In this section, the focus is on the evaluation of the errors
$\Delta z_k$ on the locations of the zeros of $H_P(z)$ due to the
approximations of the coefficients $h_P(n)$ in the polyphase
filter $H_P(e^{j\omega})$. First of all, notice that $H_P(z)$
places its $3\cdot D_1-3$ zeros in the following $z$ locations:
\[
z_k=\left\{
\begin{array}{ll}
e^{+j2\pi \frac{k}{D_1}},&~\forall k=1,\ldots,\left\lfloor
\frac{D_1}{2}\right\rfloor\\
e^{-j2\pi \frac{k}{D_1}},&~\forall k=1,\ldots,\left\lfloor
\frac{D_1}{2}\right\rfloor-1\\
e^{+j2\pi\left(\frac{k}{D_1}+ q f_c\right)}, &~\forall
k=1,\ldots,\left\lfloor \frac{D_1}{2}\right\rfloor -1\\
e^{-j2\pi\left(\frac{k}{D_1}+ q f_c\right)}, &~\forall
k=1,\ldots,\left\lfloor \frac{D_1}{2}\right\rfloor -1\\
e^{\pm j2\pi\left(\frac{k}{D_1}- q f_c\right)}, &~\forall
k=1,\ldots,\left\lfloor \frac{D_1}{2}\right\rfloor\\
\end{array}\right.
\]
The $z$-transfer function $H_P(z)$
in~(\ref{third_order_gcf3_2_poly_dec}) can be expressed in a form
emphasizing its zeros:
\begin{equation}\label{zero_displacements}
H_P(z)=\prod_{k=1}^{L-1}\left(1-z_k z^{-1}\right)
\end{equation}
whereby $L=3\cdot D_1-2$. Due to the approximation of the
coefficients $h_P(n)$, the set of zeros becomes
$\{\widetilde{z}_k=z_k+\Delta z_k,~\forall k=1,\ldots,L-1\}$:
\[
\widetilde{H}_P(z)=\prod_{k=1}^{L-1}\left(1-\widetilde{z}_k
z^{-1}\right)
\]
Zero displacements $\Delta z_k$ can be related to the coefficient
approximations $\Delta h_P$ as follows:
\begin{equation}\label{zero_displacements_1}
\Delta z_i=\sum_{\eta=1}^{L}\frac{\partial z_i}{\partial
h_P(\eta)}\Delta h_P(\eta)
\end{equation}
since there are $L$ coefficients $h_P(n)$ with $n\in [0,L-1]$.
\figuragrossa{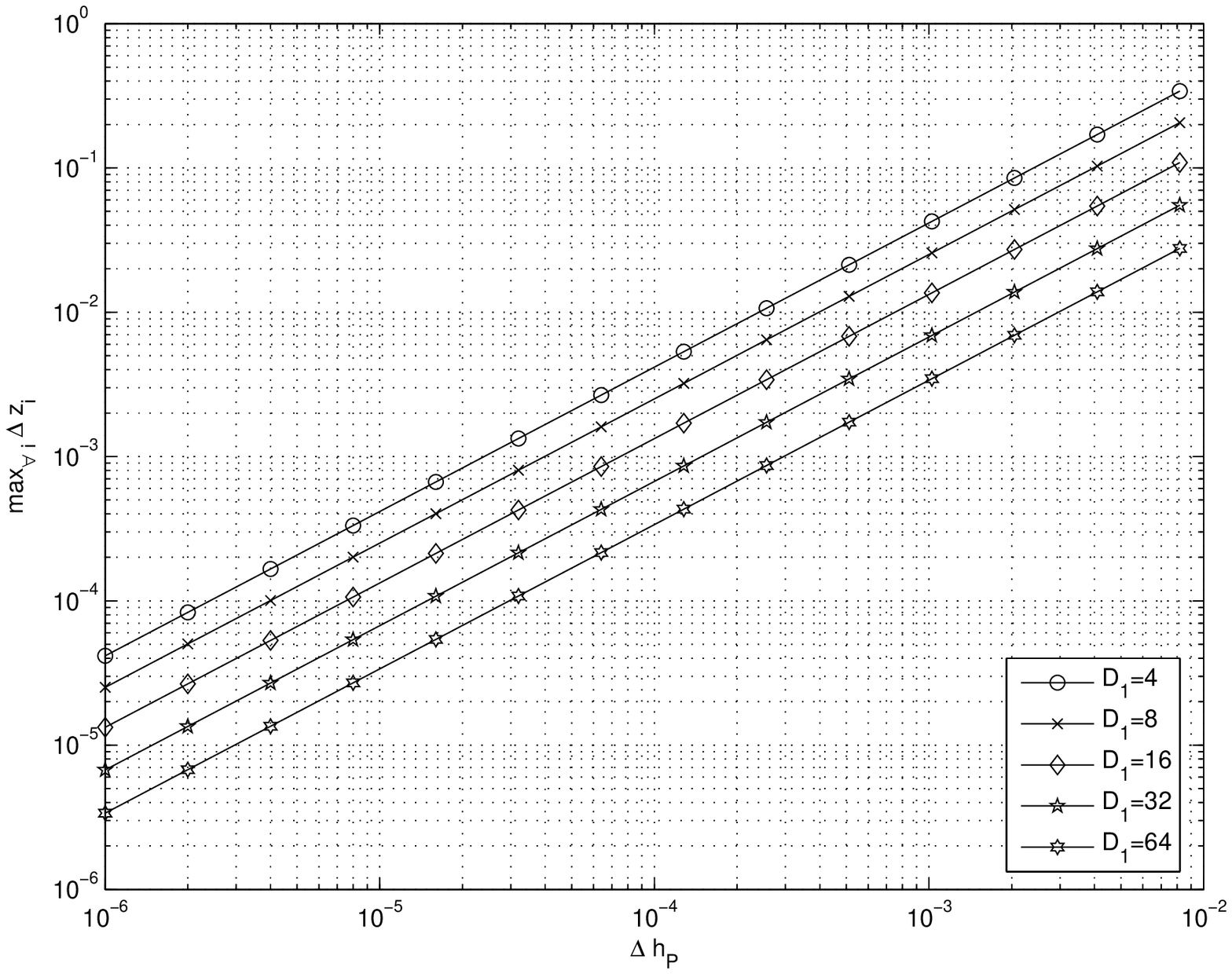}{Maximum zero displacement
over all the set of coefficients $h_P(n)$ as a function of the
approximation error $\Delta h_P$. The results assume the same
error $\Delta h_P$ for all the coefficients $h_P(n)$. Parameters
are as follows: $D_2=4$, and $\nu=4$.}{zero_displac_hp_z}

\noindent Upon noting that:
\[
\frac{\partial H_P(z)}{\partial
h_P(\eta)}_{|_{z=z_i}}=\frac{\partial H_P(z)}{\partial
z}_{|_{z=z_i}}\cdot \frac{\partial z_i}{\partial h_P(\eta)}
\]
it is straightforward to obtain:
\[
\frac{\partial H_P(z)}{\partial
h_P(\eta)}_{|_{z=z_i}}=-z^{-\eta}_{|_{z=z_i}}=-z_i^{-\eta}
\]
by employing~(\ref{third_order_gcf3_2_poly_dec}), and
\[
\frac{\partial H_P(z)}{\partial
z}_{|_{z=z_i}}=\left\{\sum_{k=1}^{L-1}\frac{z_k}{z^2}\prod_{l=1,l\ne
k}^{L-1}\left(1-z_l z^{-1}\right)\right\}_{|_{z=z_i}}
\]
by deriving~(\ref{zero_displacements}) with respect to $z$.
Finally, after some algebra~(\ref{zero_displacements_1}) can be
rewritten as follows:
\begin{equation}\label{zero_displacements_2}
\Delta z_i=-\sum_{\eta=1}^{L} \frac{z_i^{-\eta}\Delta
h_P(\eta)}{\left\{\sum_{k=1}^{L-1}\frac{z_k}{z^2}\prod_{l=1,l\ne
k}^{L-1}\left(1-z_l z^{-1}\right)\right\}_{|_{z=z_i}}}
\end{equation}
which is the displacement of the $i$-th zero $z_i$ due to the
approximations of all coefficients $h_P(n)$ belonging to the
polyphase filter decimating by $D_1$.

Fig.~\ref{zero_displac_hp_z} shows the maximum $\Delta z_i$ over
all the set of zeros indexed by $i=1,\ldots,L$, as a function of
the approximation error $\Delta h_P(\eta)=\Delta h_P$, assumed to
be the same for all coefficients, for various values of $D_1$,
$D_2=4$, and $\nu=4$. Curves shown in the figure can be considered
as the worst case zero displacement due to the approximations of
coefficients $h_P(n)$.
\subsection{Displacements of zeros belonging to $H_N(z)$}
\figuragrossa{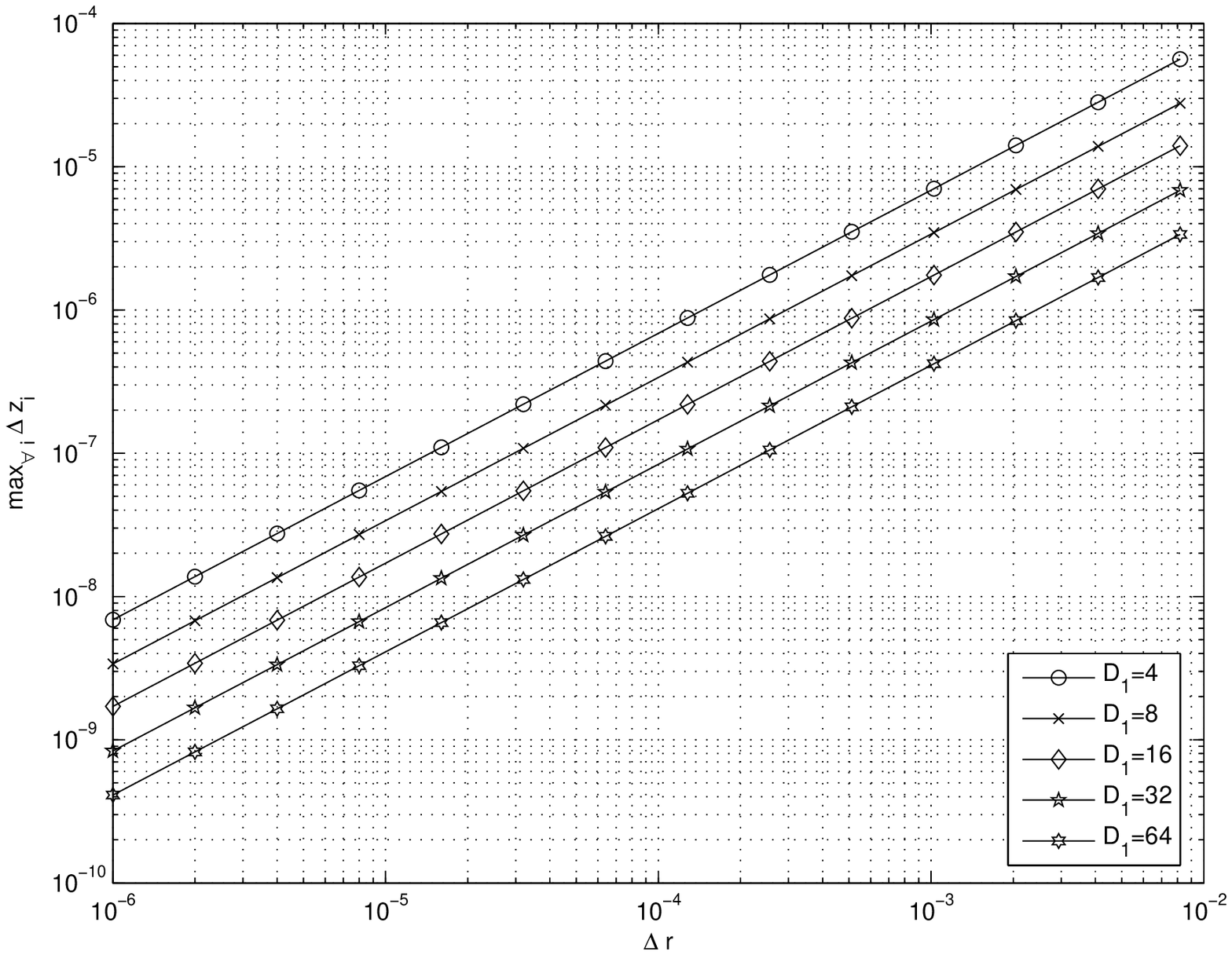}{Maximum zero displacement
over all the set of coefficients $r_u$ as a function of the
approximation error $\Delta r$. Curves assume the same error
$\Delta r$ for all the multipliers $r_u$. Parameters are as
follows: $D_2=4$, and $\nu=4$.}{zero_displac_Hn_z}
In this section, the focus is on the evaluation of the
displacements $\Delta z_k$ of the zeros in $H_N(z)$ due to the
approximation of the coefficients $r_u$ belonging to $H_N(z)$.
Error $\Delta z_i$ on the $i$-th zero can be related to the
approximation errors of multipliers $r_u$s as follows:
\begin{equation}\label{zero_displacements_1_ru}
\Delta z_i=\sum_{u=p_p+1}^{p-1}\frac{\partial z_i}{\partial
r_u}\Delta r_u
\end{equation}

\noindent Upon noting that:
\[
\frac{\partial H_N(z)}{\partial r_u}_{|_{z=z_i}}=\frac{\partial
H_N(z)}{\partial z}_{|_{z=z_i}}\cdot \frac{\partial z_i}{\partial
r_u}
\]
after some algebra on~(\ref{third_order_gcf3_2_3}), it is possible
to obtain:
\begin{equation}\label{derivata_HNz_1}
\begin{array}{ll}
\frac{\partial H_N(z)}{\partial
r_u}_{|_{z=z_i}}=\left\{\left(z^{-2^u}+z^{-2\cdot
2^u}\right)\prod_{k=p_p+1,k\ne u}^{p-1}\left[1+\right.\right.\\
\left.\left. +r_k\left(z^{-2^k}+z^{-2\cdot 2^k}\right)+z^{-3\cdot
2^k}\right]\right\}_{|_{z=z_i}}
\end{array}
\end{equation}
By deriving~(\ref{third_order_gcf3_2_3}) with respect to $z$, it
is possible to write:
\begin{equation}\label{derivata_HNz_2}
\begin{array}{ll}
\frac{\partial H_N(z)}{\partial
z}_{|_{z=z_i}}=\sum_{k=p_p+1}^{p-1}\left\{\left[r_k\left(2^k
z^{-2^k-1}\right.\right.\right.&\\
\left.\left.\left. -2\cdot 2^k z^{-2\cdot 2^k-1}\right)-3\cdot 2^k
z^{-3\cdot 2^k}\right]\cdot \right.&\\
\left.\prod_{l=p_p+1,l\ne
k}^{p-1}\left[1+r_l\left(z^{-2^l}+z^{-2\cdot
2^l}\right)+z^{-3\cdot 2^l}\right]\right\}_{|_{z=z_i}}&
\end{array}
\end{equation}
Finally,~(\ref{zero_displacements_1_ru}) can be evaluated by
substituting the ratio between~(\ref{derivata_HNz_1})
and~(\ref{derivata_HNz_2}) in place of $\frac{\partial
z_i}{\partial r_u}$.

Fig.~\ref{zero_displac_Hn_z} shows the maximum $\Delta z_i$ over
all the set of zeros belonging to $H_N(z)$, as a function of the
approximation error $\Delta r_u=\Delta r$, assumed to be the same
for all multipliers, for various values of $D_1$, $D_2=4$ and
$\nu=4$.

A quick comparison between the results shown in
Fig.s~\ref{zero_displac_hp_z} and~\ref{zero_displac_Hn_z} reveals
that the maximum zero displacement of the zeros belonging to
$H_N(z)$ is much smaller than the one experienced by zeros
belonging to $H_P(n)$. There are at least two basic reasons for
such a behaviour. First, the number of multipliers $r_u$s is very
small with respect to the number of coefficients $h_P(n)$ of the
polyphase section. Secondly, any error $\Delta r_u$ slightly
rotates the zeros belonging to the $u$-th decimation cell in
$H_N(z)$ leaving them on the unit circle. This follows from the
$z$-transfer function
$\left[1+r_u\left(z^{-1}+z^{-2}\right)+z^{-3}\right]$ of the
$u$-th decimation stage. On the other hand, approximation errors
$\Delta h_P(n)$ can also move the zeros of $H_P(z)$ outside the
unit circle in the $z$-plane worsening the $\Sigma\Delta$ QN
rejection performance of the decimation filter.

The previous analysis suggests that the most critical filter in
the partial polyphase architecture is the polyphase section
$H_P(e^{j\omega})$. In order to deduce the maximum tolerable
approximation error over the set of coefficients $\Delta h_P(n)$,
the behaviour of filter $H_P(e^{j\omega})$ around the folding
bands should be further investigated. This is the topic addressed
in the next section.
\subsection{Design Considerations}
\figuragrossa{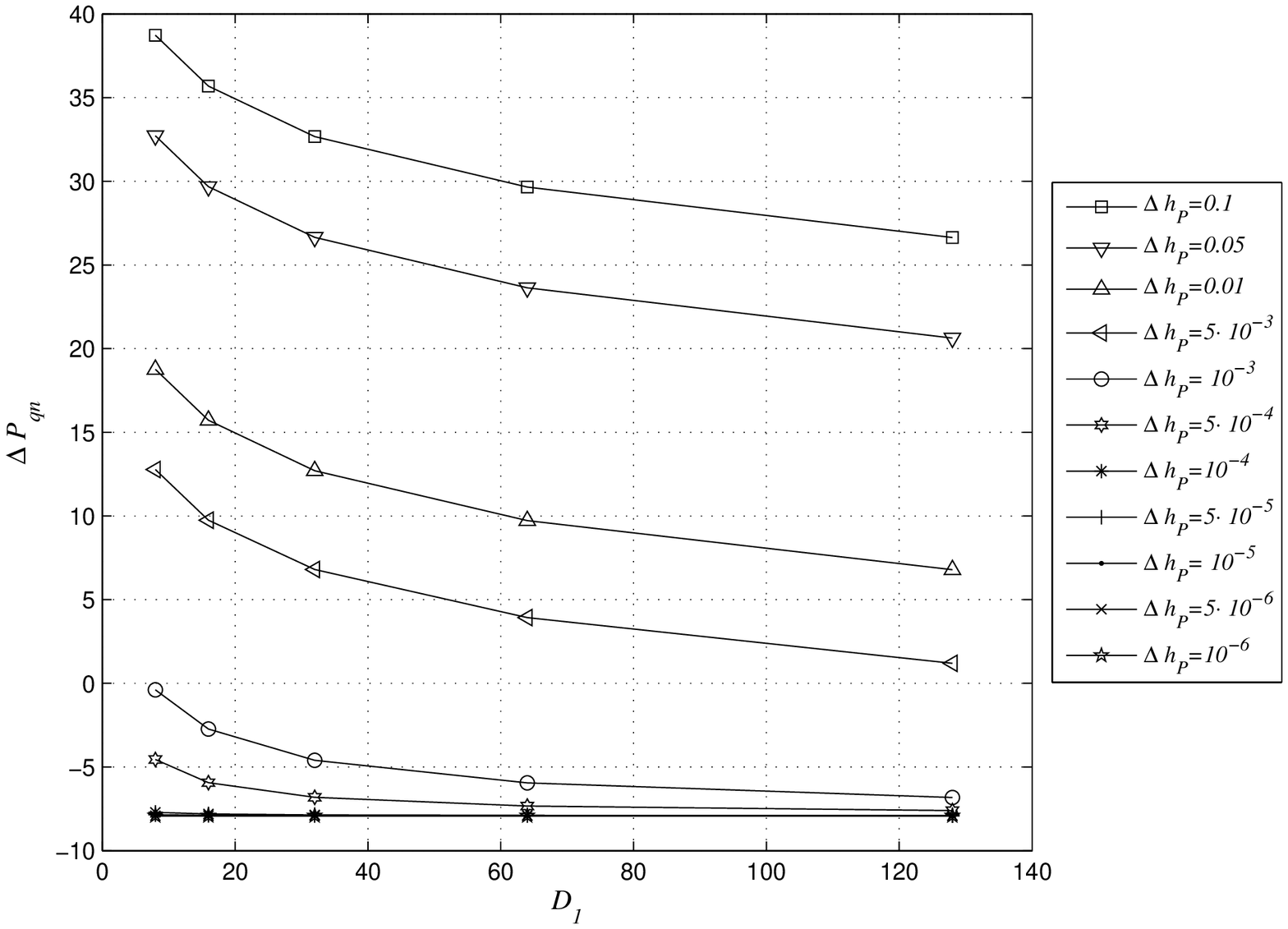}{Behaviour of the difference $\Delta
P_{qn}=\widetilde{P}_{{qn}|_{dB}}-P_{{qn}|_{dB}}$ as a function of
the decimation factor $D_1$ and for various values of the
approximation errors $\Delta h_P$ shown in the legend.}{deltapqn}
The $\Sigma\Delta$ QN power falling inside the folding bands
$[\frac{k}{D_1}-f_c,\frac{k}{D_1}+f_c],~\forall k=1,\ldots,\lfloor
\frac{D_1}{2}\rfloor$, can be defined\footnote{Note
that~(\ref{powernoise}) is valid only if the Noise Transfer
Function (NTF) of the modulator is maximally flat, i.e., it does
not contain stabilizing poles. In higher order modulators this
requires multi-bit feedback structures.} as \cite{Temes}:
\begin{equation}\label{powernoise}
P_{qn}=\sum_{k=1}^{\lfloor \frac{D_1}{2}\rfloor} \int^{
\frac{k}{D_1}+f_c}_{\frac{k}{D_1}-f_c}|H_{P}(e^{j\omega})|^2S_B(f_d)
df_d
\end{equation}
where $S_B(f_d)$, the power spectral density of the $\Sigma
\Delta$ QN, can be expressed as $S_B(f_d) = S_{e}(f_d)\cdot [2
\sin(\pi f_d)]^{2B}$. In the previous relation
$S_{e}(f_d)=\frac{\Delta^2}{12f_s}$ is the power spectral density
of the sampled noise under the hypothesis of representing the QN
as a white noise \cite{Temes}, $\Delta$ is the quantization level
of the quantizer contained in the $\Sigma\Delta$ modulator
\cite{Temes}, and $f_s$ is the $\Sigma\Delta$ sampling rate.

In order to quantify the $\Sigma\Delta$ QN rejection performance
of filter $\widetilde{H}_P(z)$ embedding the approximated
multipliers $\widetilde{h}_P(n)=h_P(n)+\Delta h_P(n)$, with
respect to a classical $3$rd order comb filter, the following
performance metric, $\Delta P_{qn}$, can be evaluated:
\begin{eqnarray}\label{gainG}
\Delta P_{qn}&=\frac{\sum_{k=1}^{\lfloor \frac{D_1}{2}\rfloor}
\int^{
\frac{k}{D}+f_c}_{\frac{k}{D}-f_c}|\widetilde{H}_{P}(f_d)|^2S_B(f_d)
df_d} {\sum_{k=1}^{\lfloor \frac{D_1}{2}\rfloor} \int^{
\frac{k}{D}+f_c}_{\frac{k}{D}-f_c}|H_{C_3}(f_d)|^2S_B(f_d) df_d}&
\end{eqnarray}
The behaviour of $\Delta P_{qn}$-[dB] as a function of $D_1$ is
shown in Fig.~\ref{deltapqn}, for various values of $\Delta h_P$,
i.e., the approximation error which is assumed to be the same for
all the coefficients $h_P(n)$. Results in Fig.~\ref{deltapqn} show
that $\Sigma\Delta$ QN rejection improvements can still be
achieved upon approximating each coefficient $h_P(n)$ within an
error less or equal to $10^{-3}$.

This in turn suggests that it is possible to optimize each
multiplier $h_P(n)$ in the polyphase filter $H_P(z)$ by
approximating it as a power-of-2 (PO2) coefficient with an
approximation error of $10^{-3}$ without affecting the
$\Sigma\Delta$ QN rejection performance of the polyphase filter
around the folding bands for any $D_1\ge 32$.

Fig.~\ref{frequency_response} shows the behaviours of both the
frequency response of filter $H^*_{GCF_3}(e^{j\omega})$ employing
approximated coefficients $\widetilde{h}_P(n)=h_P(n)+\Delta
h_P(n)$, and the frequency response of filter
$H_{GCF_3}(e^{j\omega})$ embedding the real coefficients $h_P(n)$,
for $D=D_1=64$, $\nu=4$, and $\Delta h_P(n)=\Delta
h_P=10^{-3},~\forall n$. This is the most critical case in which a
full polyphase architecture is employed. The lower subplot shows
the local behaviour of both frequency responses around some
folding bands. Notice that both curves are superimposed, even
though $H^*_{GCF_3}(e^{j\omega})$ employs coefficients
approximated with an error equal to $10^{-3}$.

The approximation of the multipliers $h_P(n)$ with PO2
coefficients is beyond the scope of this work. Nevertheless, many
excellent works have been proposed in literature for obtaining the
best PO2 coefficient approximation within a predefined error. We
invite the interested readers to refer to papers
\cite{chinglim}-\cite{lilim}.
\figuragrossa{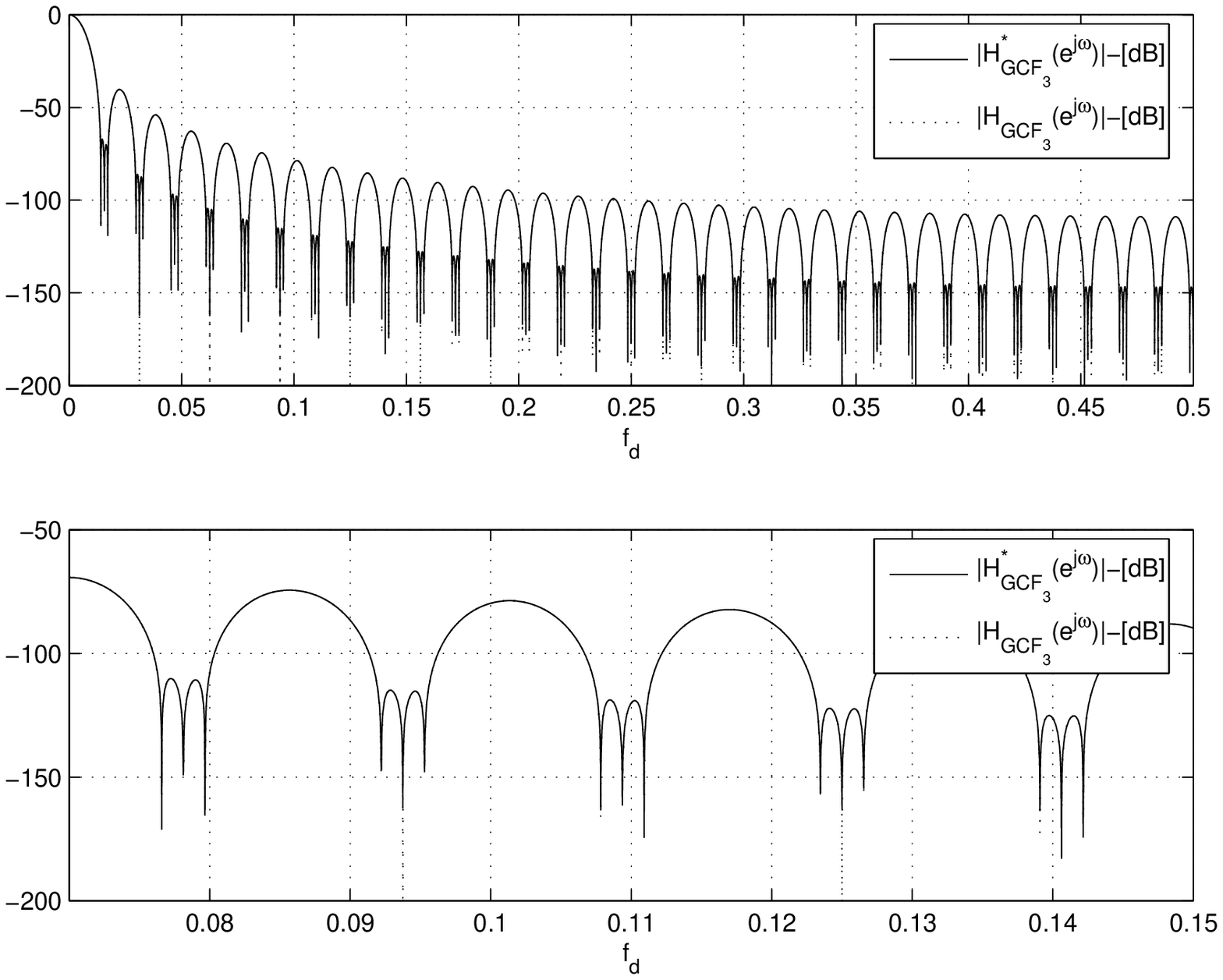}{Modulo of the frequency
response of filter $H^*_{GCF_3}(e^{j\omega})$ employing
approximated coefficients $\widetilde{h}_P(n)=h_P(n)+\Delta
h_P(n)$ (continuous curve), and frequency response of filter
$H_{GCF_3}(e^{j\omega})$ embedding the real coefficients $h_P(n)$,
for $D=D_1=64$, $\nu=4$, and $\Delta h_P(n)=\Delta
h_P=10^{-3},~\forall n$. Lower subplot shows the behaviours of
such functions around five folding bands.}{frequency_response}

Let us summarize the main design considerations deduced from the
proposed sensitivity analysis of partial polyphase GCF filters.
\begin{itemize}
    \item Both pass-band droop and selectivity performance of GCF
    filters are mainly unaffected by the approximation of the
    multipliers embedded in the decimation filter. Practically
    speaking, this means that such performance are the one already
    deduced in the companion paper~\cite{laddomada_gcf}.
    \item The most critical section in the partial polyphase
    decomposition is the polyphase filter $H_P(z)$.
    \item Upon approximating each multiplier of the polyphase
    section with an approximation error less or equal to
    $10^{-3}$, it is possible to preserve the $\Sigma\Delta$ QN
    rejection performance of GCF filters with respect to
    classical, equal-order comb filters, for a wide range of decimation factors $D_1$ as noted in the
    abscissa of Fig.~\ref{deltapqn}.
\end{itemize}
%
%
%
%
%
\section{Conclusions}
\label{conclusions}
This paper focused on the design of generalized comb filters by
proposing a novel partial polyphase architecture with the aim to
reduce the data rate after the $\Sigma \Delta $ A/D conversion. We
proposed a mathematical framework in order to analyze both the
sensitivity of the frequency response and the displacements of the
zeros in the filter transfer functions due to the quantization of
the multipliers embedded in the proposed filters.

We also derived the impulse response of a $3$rd order sample GCF
filter, that we used as a reference scheme throughout the paper.
%
%

%
%
\section*{Appendix}
In this Appendix, we derive the impulse response
$h_{GCF_3}(n),~\forall n=0,\ldots,3D-3,$ of a $3$rd-order GCF
filter decimating by $D$. The proof relies on repeated
applications of the inverse $z$-transform on the product of two
analytical functions $X_1(z)$ and $X_2(z)$ related to two
discrete-time sequences, $x_1(n)$ and $x_2(n)$:
\begin{equation}\label{inv_z_trans}
\begin{array}{ll}
Z^{-1}\left[X_1(z)X_2(z)\right]=x_1(n)\star
x_2(n)=&\\
=\sum_{k=-\infty}^{+\infty}x_1(k)x_2(n-k).&
\end{array}
\end{equation}
First of all, consider the transfer function
in~(\ref{third_order_gcf3}), and define as $X_t(z)$ the numerator
$z$-polynomial:
\[
\begin{array}{ll}
X_t(z)= \left(1-z^{-D}e^{j\alpha D}\right)
\left(1-z^{-D}e^{-j\alpha D}\right)
\left(1-z^{-D}\right)&\\
= \left[1-r z^{-D}+r z^{-2D}-z^{-3D}\right]&
\end{array}
\]
whereby $r=1+2\cos(\alpha D)$.
The discrete-time, causal sequence with $z$-transfer function
$X_t(z)$ can be written as follows:
\begin{equation}\label{inv_z_trans_X_tz}
x_t(n)= \delta(n)-r\delta(n-D)+r\delta(n-2D)-\delta(n-3D)
\end{equation}
whereby $\delta(n)$ is the discrete-time unit impulse centered in
$n=0$.

Let us define the following pairs of transfer functions along with
the respective discrete-time sequences \cite{antoniou}:
\[
\begin{array}{lll}
Y_1(z)=\frac{1}{1-z^{-1}}&\longleftrightarrow & y_1(n)=u(n)\\
Y_2(z)=\frac{1}{1-z^{-1}e^{-j\alpha}}&\longleftrightarrow & y_2(n)=e^{-j\alpha}u(n)\\
Y_3(z)=\frac{1}{1-z^{-1}e^{j\alpha}}&\longleftrightarrow &
y_3(n)=e^{+j\alpha}u(n)
\end{array}
\]
whereby $u(n)$ is the discrete-time unitary-step sequence.

With the setup above, $H_{GCF_3}(z)$ in~(\ref{third_order_gcf3})
can be rewritten as follows:
\[
H_{GCF_3}(z)=X_t(z)\cdot Y_1(z)\cdot Y_2(z) \cdot Y_3(z)
\]
Upon applying~(\ref{inv_z_trans}) to the $z$-function
$W_1(z)=X_t(z)Y_1(z)$, it is possible to obtain:
\begin{equation}\label{inv_z_trans_1}
w_1(n)=\sum_{k_1=-\infty}^{+\infty}x_t(k_1)y_1(n-k_1)
\end{equation}
Applying~(\ref{inv_z_trans}) to the $z$-function
$W_2(z)=W_1(z)Y_2(z)$, and employing~(\ref{inv_z_trans_1}), it is
possible to obtain:
\begin{equation}\label{inv_z_trans_2}
\begin{array}{ll}
w_2(n)=\sum_{k_2=-\infty}^{+\infty}w_1(k_2)y_2(n-k_2)=&\\
=\sum_{k_2=-\infty}^{+\infty}\sum_{k_1=-\infty}^{+\infty}x_t(k_1)y_1(k_2-k_1)
y_2(n-k_2)&
\end{array}
\end{equation}
Finally, applying~(\ref{inv_z_trans}) to the $z$-function
$H_{GCF_3}(z)=W_3(z)=W_2(z)Y_3(z)$, and
employing~(\ref{inv_z_trans_2}), it is possible to obtain:
%
\begin{eqnarray}\label{inv_z_trans_3}
h_{GCF_3}(n)=\sum_{k_3=-\infty}^{+\infty}w_2(k_3)y_3(n-k_3)=&\nonumber\\
=\sum_{k_3=-\infty}^{+\infty}\sum_{k_2=-\infty}^{+\infty}
\sum_{k_1=-\infty}^{+\infty}\left[x_t(k_1)y_1(k_2-k_1)\cdot\right.\nonumber&\\
\left.\cdot y_2(k_3-k_2)y_3(n-k_3)\right]&
\end{eqnarray}
%
Upon substituting the respective expressions of the sequences
$y_i(n),~\forall i=1,\ldots,3$ in~(\ref{inv_z_trans_1}), it is
possible to write:
%
\begin{eqnarray}\label{inv_z_trans_4}
h_{GCF_3}(n)=\sum_{k_3=-\infty}^{+\infty}\sum_{k_2=-\infty}^{+\infty}
\sum_{k_1=-\infty}^{+\infty}\left[x_t(k_1)u(k_2-k_1)\cdot\right.&\nonumber\\
\left.\cdot e^{-j\alpha (k_3-k_2)}u(k_3-k_2)  e^{+j\alpha
(n-k_3)}u(n-k_3)\right]&
\end{eqnarray}
%
By exploiting the definitions of the unitary-step sequences, it is
possible to observe the following relations:
\[
u(k_2-k_1)=\left\{
\begin{array}{ll}
1&~\forall k_2\ge k_1 \\
0 &~\forall k_2< k_1
\end{array}\right.
\]
\[
u(k_3-k_2)=\left\{
\begin{array}{ll}
1&~\forall k_3\ge k_2 \\
0 &~\forall k_3< k_2
\end{array}\right.
\]
\[
u(n-k_3)=\left\{
\begin{array}{ll}
1&~\forall n\ge k_3 \\
0 &~\forall n< k_3
\end{array}\right.
\]

Based on these observations, we can reduce the upper limits of the
summations in~(\ref{inv_z_trans_4}) as follows:
\begin{eqnarray}\label{inv_z_trans_5}
h_{GCF_3}(n)=\sum_{k_3=-\infty}^{n}\sum_{k_2=-\infty}^{k_3}
\sum_{k_1=-\infty}^{k_2}\left[x_t(k_1)\cdot\right. &\nonumber\\
\left.\cdot e^{-j\alpha (k_3-k_2)} e^{+j\alpha (n-k_3)}\right]
\end{eqnarray}
By observing that the sequence $x_t(k_1)$ is causal
(see~(\ref{inv_z_trans_X_tz})), i.e., $x_t(k_1)=0,~\forall k_1<0$,
it is possible to obtain:
\begin{eqnarray}\label{inv_z_trans_6}
h_{GCF_3}(n)=e^{+j\alpha n}\sum_{k_3=0}^{n}e^{-2j\alpha
k_3}\sum_{k_2=0}^{k_3}e^{j\alpha k_2} \sum_{k_1=0}^{k_2}x_t(k_1)&
\end{eqnarray}
Notice that the impulse response $h_{GCF_3}(n)$ generalizes the
one of the classical $3$rd order comb decimation filter, and, it
is composed by $3D-2$ coefficients over the time interval ranging
from $0$ to $3D-3$. Equ.~(\ref{inv_z_trans_6}) can also be used as
an on-line algorithm for generating the coefficients of the GCF
impulse response $h_{GCF_3}(n)$ by simply solving the three nested
summations for each $n=0,\ldots,3D-3$.

As a note aside, notice that by imposing $\alpha=0$
in~(\ref{inv_z_trans_6}), it is possible to obtain the impulse
response of a classical $3$rd-order comb filter.

\end{document}